\documentstyle[osa,manuscript]{revtex}
\begin{document}
\title{Second-harmonic interferometric spectroscopy \\
of the buried Si(111)-SiO$_2$ interface}
\author{A.A. Fedyanin, T.V. Dolgova, O.A. Aktsipetrov}
\address{Department of Physics, Moscow State University, 119899 Moscow, Russia}
\author{D. Schuhmacher, G. Marowsky}
\address{Laser-Laboratorium G\"ottingen, Hans-Adolf-Krebs-Weg 1, D-37077 G\"ottingen, Germany}
\maketitle

\begin{abstract}
The second-harmonic interferometric spectroscopy (SHIS) which
combines both amplitude (intensity) and phase spectra of
the second-harmonic (SH) radiation is proposed as
a new spectroscopic technique being sensitive to the type of
critical points (CP's) of combined density of states
at semiconductor surfaces. The increased sensitivity of SHIS technique is demonstrated for the buried
Si(111)-SiO$_2$ interface for SH photon energies from 3.6 eV to 5 eV
and allows to separate the resonant contributions from
$E^\prime_0/E_1$, $E_2$ and $E^\prime_1$ CP's of silicon.
\end{abstract}

Second-harmonic
generation (SHG) is inherently sensitive to surface and interface properties of
centrosymmetric materials. Recently, the spectroscopy of
the second-harmonic (SH) intensity has been proved as
a promising probe of surfaces and interfacial
layers \cite{heinz} and
intensively employed in numerous works
for oxidized \cite{daum1}, reconstructed \cite{pedersen} and
H-terminated \cite{dadap1} silicon surfaces. The resonances of
the SH intensity  are attributed in these cases to direct
interband electron
transitions.
By analogy with the spectrum of the linear
dielectric function $\varepsilon(\omega)$
the spectrum of the quadratic susceptibility
$\chi^{(2)}(2\omega)$ of such
semiconductor surfaces
could be expressed as the superposition of
several van Hove singularities
(critical points (CP's) of the combined
density of states)
$\chi^{(2)}_m(2\omega) \propto (2\omega-\omega_m+i\Gamma_m)^n$
 with threshold frequencies $\omega_m$ and
broadenings $\Gamma_m$\cite{cardona1}.
The exponent $n$ reflects the
dimensionality of CP: $n=1/2,0$ (logarithmic),$-1/2, -1$
for 3D, 2D, 1D and excitonic CP, respectively.
Although the line shapes $\chi^{(2)}(2\omega)$  are quite different
for various $n$, in most cases
a large number of adjustable parameters
makes the determination of the type of CP solely from the
SH intensity spectrum doubtful, and most of authors interpret
the SHG spectroscopy data within excitonic CP line shape\cite{daum}.

The single-beam SH interferometry traces back to
the mid-1960s \cite{chang} and is
conventionally used for the determination of the phase of the quadratic susceptibility
of adsorbate molecules and their absolute direction \cite{kemnitz}
and for separation of the SH contributions from thin films and their
substrates \cite{stolle}. Another
use of SH phase measurements is a homodyne mixing
technique to improve the signal-to-noise ratio for surface SHG probe \cite{superfine}.
The use of
external \cite{dadap} and internal \cite{aktsip1} homodynes for
dc or low-frequency modulated electric-field-induced SHG allows to measure the
in-plane spatial distribution of
the electric field vector with micron resolution or to visualize
weak nonlinear contributions. Further developments of the SH interferometry are the
frequency-domain interferometric SH
spectroscopy \cite{wilson} exploring the broad bandwidth of femtosecond laser pulses,
and the hyper-Rayleigh scattering
interferometry \cite{aktsip2} using the correlation of fluctuations in linear and
nonlinear optical properties of thin
inhomogeneous films.

In this Letter, a modification
of the SH spectroscopy - the SH interferometric spectroscopy (SHIS)
which combines both amplitude (intensity) SH spectroscopy
and SH interferometry is proposed. The combination
of the phase and amplitude SH spectra,
extracted from the SHIS data, is shown to be sensitive
to type of CP even for systems with
interfering SH contributions from close electronic
resonances. Additionally SHIS allows to avoid
the sign uncertainty of ${\rm Re}(\chi^{(2)})$
inherent in the conventional spectroscopy of the SHG intensity.
 The spectral dependence of
the phase and amplitude of the SH waves from the buried
Si(111)-SiO$_2$ interface is measured using the SHIS technique
in the spectral range
in the vicinity of silicon $E_2$ CP. In contrast to $E^\prime_0/E_1$
CP revealing the excitonic type, the family of $E_2$ CP's
of the bulk silicon demonstrates the 2D type in linear
response\cite{cardona}.  The resonant contributions
to the quadratic susceptibility from $E^\prime_0/E_1$, $E_2$ and
$E^\prime_1$ silicon CP's are extracted within the simple
phenomenological model which
accounts the complex Green's function corrections for the SH wave
generation.

The scheme of the SHIS setup is shown in Fig.1(a). The p-polarized
output of a tunable nanosecond parametric generator/amplifier laser
system (Spectra-Physics MOPO 710) operating in the interval
of 490 - 690 nm is focused onto the sample at an angle of incidence of $45^\circ$. The SH
signal is detected by a monochromator, a photomultiplier tube (PMT)
and an electronic peak-hold detector. To normalize the SH
intensity spectrum over the laser fluence and the spectral sensitivity
of the optical detection system a SHG intensity reference channel is used
with a slightly wedged z-cut quartz
plate and with the detection system identical to the one in the sample channel.
The (phase-)reference sample is chosen
(i) to be thin enough to avoid Maker fringes in the SH response during tuning
the fundamental wavelength $\lambda_\omega$, (ii) to be optical
inactive for conservation the polarization state of the fundamental radiation
while transmitting through it, (iii) to have no resonance features in the
tuning region of both the fundamental and SH
waves. Therefore the 1 mm-thick plate of fused quartz
coated with a 30 nm-thick indium tin oxide (ITO) film is chosen as
a reference. The SH interferogram is obtained by translating the reference
along the fundamental laser beam
varying the distance {\it l} between the reference and the sample. The SH signals from
the reference, $I^{2\omega}_r$, and from the sample, $I^{2\omega}_s$, are
monitored separately by inserting appropriate filters (yellow
or UV, respectively) between the reference and the sample.
$I^{2\omega}_r$ is adjusted with the angle of incidence of the fundamental beam
at the ITO phase reference.
The detected SH intensity $I^{2\omega}$ is the result
of interference of the SH waves from the reference,
$E^{2\omega}_r$, and from the sample, $E^{2\omega}_s$:
\begin{equation}
I^{2 \omega}={c \over {8\pi}} | E^{2\omega}_r (l)+ E^{2\omega}_s
| ^2= I^{2\omega}_r(l)+ I^{2\omega}_s +2 \alpha
\sqrt{I^{2\omega}_r(l)I^{2\omega}_s } \cos \left(2\pi\frac{l}{L} + \Phi_{rs} \right),
\label{eq1}
\end{equation}
where $L=\lambda_\omega (2\Delta n)^{-1}$ is the period of SH
interferogram with $\Delta n=n_{2\omega}- n_\omega$ describing the
air dispersion, and $\alpha < 1$ indicates the laser coherence.
The position-dependent phase shift $2\pi l/L$ between $
E^{2\omega}_r$ and $E^{2\omega}_s$ comes from the different
refractive indices of air for the fundamental and SH waves. The
spectral dependences of $\chi^{(2)}$ of the reference and the
sample as well as the complex Green's function corrections for the
SH wave produce a position-independent phase shift
$\Phi_{rs}(\lambda_\omega, \lambda_{2\omega}) $. The dependence
$I^{2\omega}_r(l)$ is described by the conventional formula for
focused Gaussian beams. The spectrum of the phase of the SH wave
from the ITO film, $\Phi_r\equiv{\rm Arg}(E^{2\omega}_r)$, is
measured using the 1 mm-thick backside-immersed y-cut quartz as a
sample since the phase of the SH wave from the quartz surface is
spectrally independent in the whole used spectral region. The
$\Phi_r$ spectrum of the ITO film appears to be a constant within
the error bars and $I^{2\omega}_r$ gradually increases with
decreasing $\lambda_\omega$ without any resonance features.

The samples are natively oxidized p-doped Si(111)
wafers with resistivity of $10\Omega\cdot cm$. SHIS has been performed at the maximum of
the azimuthal SH rotational anisotropy for the p-in, p-out polarization combination
of the fundamental and SH waves.
Figure 1(b) shows typical SH interference patterns measured for different $\lambda_\omega$.
The spectral dependence of the period $L$
due to the air dispersion and clear changes in the contrast of
the patterns due to the distance dependence of $I^{2\omega}_r$ in the
focused laser beam are seen. The fit of the set of SH
interference patterns by Eq.(\ref{eq1}) with $\Phi_{rs}$, $I^{2\omega}_s$,
$L$, and
$\alpha$ as adjustable parameters leads to the spectra of
$\Phi_{rs}$, $I^{2\omega}_s$
and $L$ shown in Fig. 2(a) and 2(b) and in the inset
of Fig. 1, respectively.
To emphasize the spectral features of $I^{2\omega}_s(2\omega)$,
we combine the fitted intensity spectrum with the $I^{2\omega}_s(2\omega)$
dependence measured directly with a fine resolution in SH photon energy.
$\Phi_{rs}$
increases approximately by 1.2 radians within the interval
of 4.2-4.6 eV  and decreases outside
this energy region.
A small, but reliable,
non-monotonic feature is seen at 3.8 - 4.0 eV. The $I^{2\omega}_s$
spectrum has pronounced peaks centered approximately
at 3.9 eV and 4.3 eV. The position of the 4.3 eV resonance is close
to $E_2$ CP and we associate the observed features of SH
phase and intensity spectra for energies between 4.1 and 4.6 eV with
direct interband electron transitions at $E_2$ CP of
Si \cite{cardona}.

The relative phase $\Phi_{rs}$ measured in SHIS is given by:
\begin{equation}
\Phi_{rs}=\Phi_s-(\Phi_r+{\rm Arg}(R_{2\omega})),
\label{eq2}
\end{equation}
where $ R_{2\omega}$ is the Fresnel reflection factor of the p-polarized SH radiation
from the Si-SiO$_2$ interface.
The phase $\Phi_s\equiv {\rm Arg}(E_s^{2\omega})$
originates from both complex surface $\chi^{(2),S}$
and bulk quadrupole $\chi^{(2),BQ}$ quadratic
susceptibilities
as well as from Green's
function corrections \cite{sionnest}:
\begin{equation}
E_s^{2\omega}=G_\parallel\chi^{(2)}_\parallel+G_\perp\chi^{(2)}_\perp,
\label{eq3}
\end{equation}
where $ G_\parallel$ and $G_\perp$ are the Green's function corrections
for the generation and the propagation of in-plane and
normal components of $E^{2\omega}_s$,
and $\chi^{(2)}_\parallel$ and $\chi^{(2)}_\perp$ are the corresponding
effective components of $\chi^{(2)}$.
$\chi^{(2)}_\parallel$ and $\chi^{(2)}_\perp$ are
the linear combinations of $\chi^{(2),S}$ and
$\chi^{(2),BQ}$ components with nonresonant coefficients
depending only on the fundamental wavevector
and taking into
account the geometry of the nonlinear interaction.
This allows to consider the spectral dependences of
$\chi^{(2)}_\parallel$ and $\chi^{(2)}_\perp$ as a superposition of
two-photon resonances for different CP's:
\begin{equation}
\chi^{(2)}_{\alpha}(2\omega)=B-
\sum\limits_{m} {f_m^\alpha\exp(i\phi_m^\alpha)(2\omega-\omega_m+i\Gamma_m)^n},
\label{eq4}
\end{equation}
where $\alpha=\perp,\parallel$, and $m$ numerates the CP
resonances. The oscillator strengths $ f_m^\alpha $ are
supposed to be real numbers. For the sake of simplicity, a slight
spectral dependence of the term $B$ including the Si resonances
with threshold energies below 1.5 eV \cite{pedersen} is neglected,
and $\phi_m^\alpha$ are integer multiples of $\pi/2$ defining the
type of CP. The solid lines in Fig.2 show the fit of $\Phi_{rs}$
and $I^{2\omega}_s$ spectra by Eq.(\ref{eq2}) and
$|E_s^{2\omega}|^2$ from Eq.(\ref{eq3}), respectively, with the Si
dispersion data from Ref.[\cite{aspnes2}] and expressions for $
G_\parallel$ and $G_\perp$ from Ref.[\cite{aktsip3}]. Five resonant
contributions are included into the fit. The first resonance,
centered at $\omega_1=3.45 $ eV, has excitonic line shape ($n=-1,
\phi_1=0$) and corresponds to the direct electron transitions at
$E^\prime_0/E_1$ CP. The second resonance at 3.97 eV with 1D
maximum line shape ($n=-1/2, \phi_2=0$) has no equivalent in the
band structure of crystalline bulk silicon. However, a resonance
in the close energy interval has been recently observed at the
Si(001)-SiO$_2$ interface \cite{daum} and could be associated with
transition in Si atoms located at the interface with reduced
lattice symmetry. The strong resonant features in the vicinity of
4.3 eV are formed by interference of two resonances centered at
$\omega_3=4.12$ eV and at $\omega_4=4.34$ eV  with almost equal
amplitudes ($f_4\approx0.9f_3$). It is mostly reasonable to
attribute these peaks to transitions at $E_2(X)$ and $E_2(\Sigma)$
CP's. These resonances are fitted with 2D minimum ($\phi_3=0$) and
2D maximum ($\phi_4=\pi$) line shapes ($\chi^{(2)}_m\propto
\ln(2\omega-\omega_m+i\Gamma_m)$) by analogy with the linear
case\cite{cardona}. Note, that $\omega_3$ and $\omega_4$ are
approximately 0.1 eV red-shifted from resonances of linear
$\chi^{(1)}$. This allows to interpret also 4.12 eV-resonance as a
contribution from $E_0$ CP, which is normally very weak in the
linear response. The best representation of the data is obtained
with a 2D minimum line shape of the last resonance, centered at
5.15 eV, which can be associated with electron transitions located
near $E^\prime_1$ CP\cite{cardona}. The error bars for the central
frequencies are approximately 0.03 eV and mostly attributed to the
relative weight of $\chi^{(2)}_\parallel$ and $\chi^{(2)}_\perp$
being unresolvable from our data. Excitonic line shapes for all
resonances (dotted lines in Fig.2) fit the $I^{2\omega}_s$
spectrum with almost the same quality as the CP model, but fit the
$\Phi_{rs}$ spectrum obviously worse.

Summarizing, the general scheme of the second-harmonic
interferometric spectroscopy is presented.
The phase and amplitude of the SH wave from
the buried Si(111)-SiO$_2$ interface are measured simultaneously
 using the SHIS
technique in the interval
of SH photon energies from 3.6 eV to 5 eV. The contributions
of interband transitions located at $E^\prime_0/E_1$, $E_2$ and
$E^\prime_1$ Si critical points are separated and sensitivity
of SHIS to CP line shapes is shown.

This work was supported by the Russian Foundation for Basic Research (RFBR)
and Deutsche Forschungsgemeinschaft (DFG):
RFBR grant 98-02-04092, DFG grants 436 RUS 113/439/0
and MA 610/20-1, RFBR grant 00-02-16253,
special RFBR grant for Leading Russian Science Schools 00-15-96555;
NATO Grant PST.CLG975264, Russian Federal Program
"Center of Fundamental Optics and Spectroscopy", and
Program of Russian Ministry of Science and Technology
"Physics of Solid State Nanostructures".

\newpage
FIGURES

Fig. 1. Panel a: Experimental setup for the SH interferometric spectroscopy.
BS, beam splitter; GG and UG, yellow and UV filters, respectively.
Panel b: Raw SH interferograms for different SH energies. Solid curves: The dependences given
by Eq.(\ref{eq1}). Inset: The spectral dependence of the period $L$ of
the SH interferograms and its fit using a phenomenological expression
for air dispersion. Open circles indicate the periods for the curves at the main panel.

Fig. 2. Spectrum of the SH phase $\Phi_{rs}$ (panel a)
and SH intensity $I_s^{2\omega}$ (panel b).
Solid curves are fits to the data within the model of CP line shapes.
Dotted lines are fits with excitonic line shapes for all the
resonances.

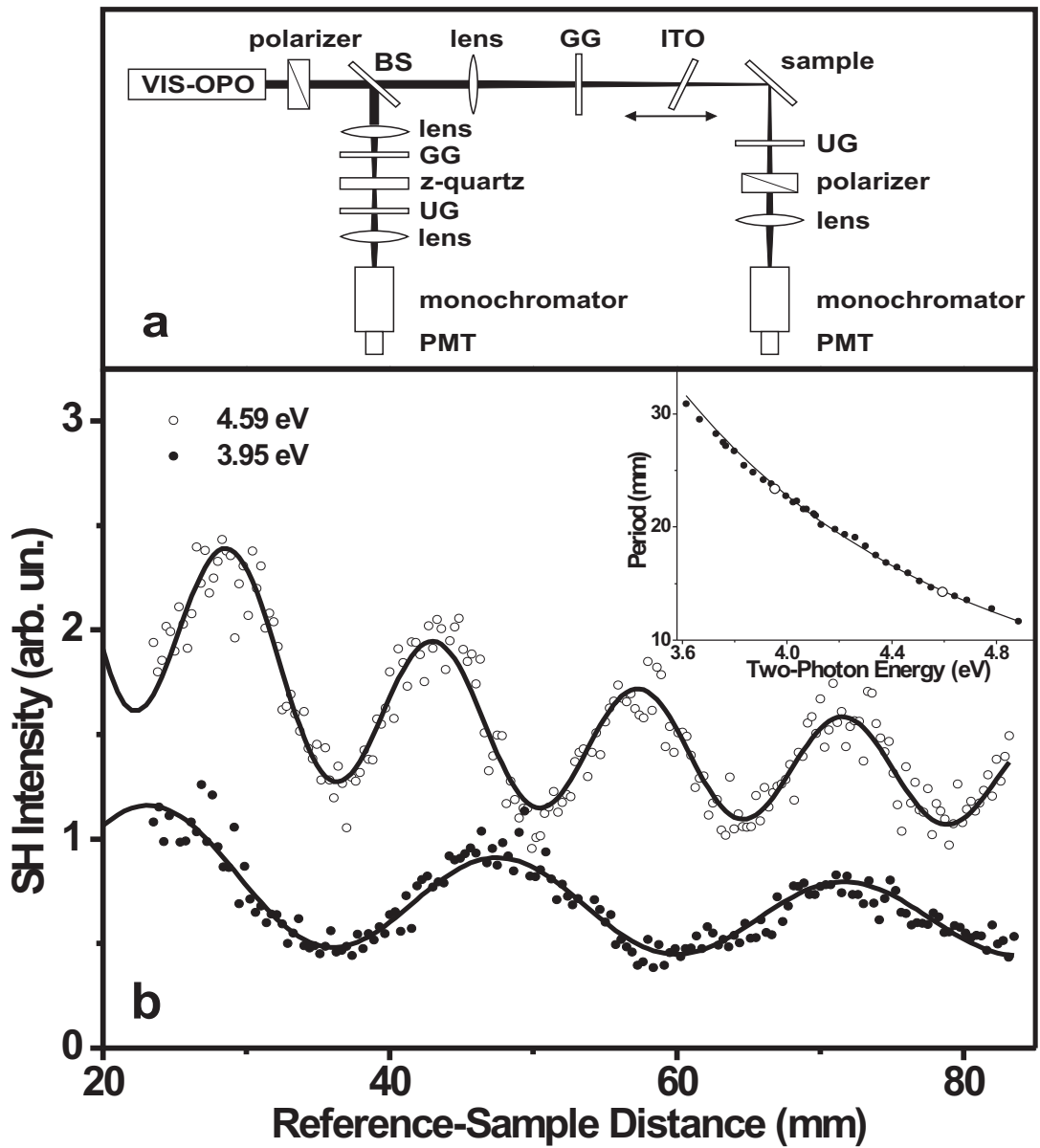
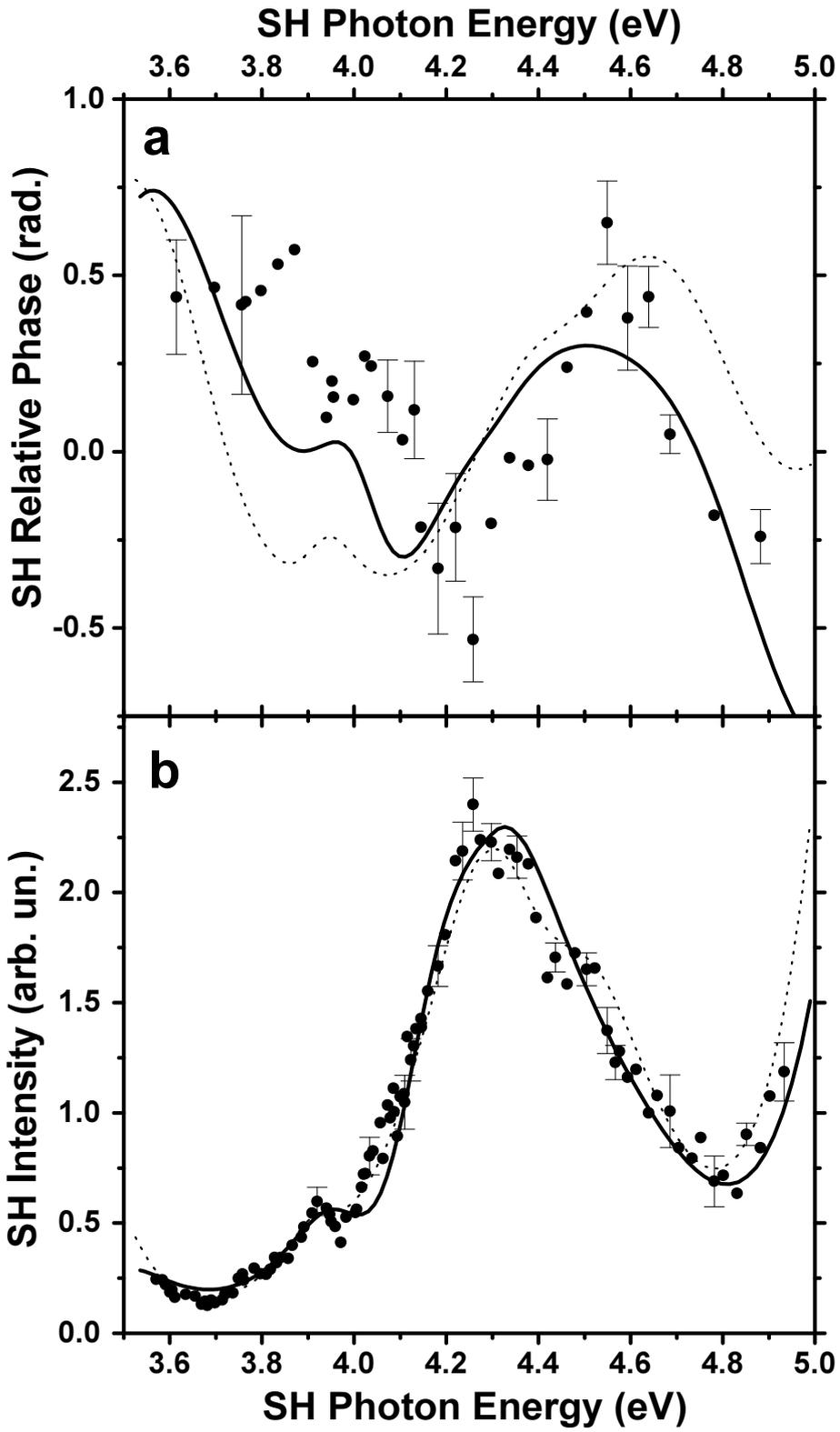
\end{document}